\documentstyle[aps,twocolumn,eqsecnum]{revtex}

\begin{document}
\author{Itzhak Bars\thanks{%
Research supported in part by the DOE Grant No. DE-FG03-84ER-40168. }\medskip}
\address{Department of Physics and Astronomy\\
University of Southern California\\
Los Angeles, CA 90089-0484}
\title{{\small \noindent hep-th/9503228 \hfill USC-95/HEP-B2 } \bigskip\\
Stringy evidence for D=11 structure \\
in strongly coupled type-IIA superstring}
\date{March 31, 1995}
\maketitle

\begin{abstract}
\centerline{\bf Abstract}\smallskip

Witten proposed that the low energy physics of strongly coupled D=10
type-IIA superstring may be described by D=11 supergravity. To explore the
stringy aspects of the underlying theory we examine the stringy massive
states. We propose a systematic formula for identifying non-perturbative
states in D=10 type-IIA superstring theory, such that, together with the
elementary excited string states, they form D=11 supersymmetric multiplets,
in SO(10) representations. This provides hints for the construction of a
conjectured weakly coupled D=11 theory that is dual to the strongly coupled
D=10 type IIA superstring.
\end{abstract}

Pacs: {11.17.+y,02.40.+m,04.20.Jb}


\section{Introduction}

Recently several proposals have been made about duality connections between
various string theories in several dimensions. These seem to provide a
handle on the behavior of strongly interacting string theories in various
dimensions. In this paper we will concentrate on the type IIA D=10
superstring in the strong coupling limit. In a recent paper Witten \cite
{witten} made the remarkable suggestion that 11 dimensional supergravity,
including all the states that come from compactifying the 11th dimension,
represents the {\it low energy sector} of the {\it strongly coupled} type
IIA D=10 supertring. He suggested that non-perturbative black hole or
monopole type states that are alleged to arise in the strongly coupled D=10
string theory are precisely the massive Kaluza-Klein states of D=11
supergravity. Also, as was known for a long time, the original massless
string states are the massless D=11 supergravity states. He based his
reasoning on U-duality and hints provided in related previous work \cite
{hulltowns}.

I first give a short summary of Witten's idea in a form that will suggest
generalizations. The argument relies on extending the D=10, N=2 super
Poincar\'e algebra with a central extension. There is mounting evidence that
the D=10 type-IIA superalgebra develops a central extension $Z$
non-perturbatively
\begin{equation}
\left\{ Q_\alpha ^i,Q_\beta ^j\right\} =p^\mu \left( C\gamma _\mu \right)
_{\alpha \beta }+\varepsilon ^{ij}C_{\alpha \beta }\,Z,  \label{susy10}
\end{equation}
and that there are states that carry non-trivial values of $Z.$ Witten
argued that the values of $Z$ should be given by $Z=cW/\lambda $ where $c$
is a pure constant, $W$ is quantized in terms of a unit, $W=nW_0,$ and $%
\lambda $ is the coupling constant of the interacting string. It is useful
for our arguments later to regard the central extension $Z$ as the $11th$
component of momentum for an 11-dimensional superalgebra
\begin{equation}
\left\{ Q_A,Q_B\right\} =P\cdot \left( C\Gamma \right) _{AB}  \label{susy11}
\end{equation}
where $A,B=1,2,\cdots 32,$ and $P$ is a momentum in 11-dimensions $P=$($%
p^\mu ,Z).$ Then (\ref{susy10},\ref{susy11}) are the same superalgebra. It
is useful to rewrite the algebra in the rest frame $P\rightarrow P^{\prime
}=(M_{11},\vec 0,0)$ where $M_{11}^2=M^2-Z^2.$ If $M_{11}$ vanishes, it is
not possible to go to the rest frame. It is known that when the mass $M_{11}$
vanishes the supersymmetry representations are ``short'' and contain $2^7$
bosons plus $2^7$ fermions. These states have $M=\left| Z\right| =c\left|
W\right| /\lambda $ and they are called the ``BPS saturated states'' in \cite
{witten}. When $M_{11}\neq 0$ the supersymmetry representations are
``long'', and contain the seeds of an SO(10) little group, as will be
discussed below.

For zero charge $W=0$ the ``BPS saturated states'' are precisely the
massless states of the string theory, and their quantum numbers are
determined by the $2_B^7+2_F^7$ short multiplet. Since the little group for
massless states in D=11 is SO(9), it is natural to expect that the massless
string states should be classified by SO(9). So, even though in D=10 the
little group for massless states is only SO(8), this argument demands that
the SO(8) states have just the property to fit into SO(9) multiplets.
Indeed, the massless states which come from the Ramond-Ramond vacuum sector $%
|$vac$>_L|$vac$>_R$, are classified under SO(8) as $\left( 8_v+8_{-}\right)
_L\times \left( 8_v+8_{+}\right) _R,$ where $8_v,8_{\pm }$ are the vector
and spinor representations of SO(8). This product yields the following
representations for the massless bosons and fermions
\begin{eqnarray}
SO(8) &:&\left( 1+8+28+35+56\right) _B \\
&&+\left[ 8_{+}+8_{-}+56_{+}+56_{-}\right] _F\;,  \nonumber
\end{eqnarray}
and, as is well known they do form the SO(9) multiplets
\begin{eqnarray}
SO(9) &:&\,\quad 2_B^7=(44+84)_B,\quad 2_F^7=128_F  \label{sugra} \\
\phi _{(IJ)} &=&44,\quad \phi _{[IJK]}=84,\quad \psi _{\alpha I}=128
\nonumber
\end{eqnarray}
which are interpreted as the graviton, antisymmetric tensor, and gravitino
of D=11 supergravity.

For non-zero charge $W\neq 0$ the BPS saturated states are massive $%
M=c\left| W\right| /\lambda ,$ but as argued above, they are classified in
the same short multiplets. Witten interpreted them as the Kaluza-Klein
excitations of the graviton, antisymmetric tensor and the gravitino, coming
from compactifying the 11th dimension of D=11 supergravity. Furthermore he
suggested that their interactions at low energies are given precisely by the
full theory of D=11 supergravity.

These arguments suggest that there may be some richer, stringy, D=11-like
theory that is dual to D=10 type-IIA superstring, whose spectrum and
interactions may be studied in either the language of the D=11 theory (if
one could guess it) or directly in the original D=10 theory. If this is true
we should be able to see more evidence of D=11 by studying the type IIA
theory. In particular it should be possible to see that the stringy spectrum
of the D=10 string theory exhibits some D=11 property. Indeed we have known
for quite some time that the first stringy massive level of the type IIA
theory does exhibit a D=11 dimensional structure\cite{ibmembrane}. Namely
the states fill the long multiplet $2_B^{15}+2_F^{15}$ of D=11
supersymmetry, and they are classified in SO(10) representations, i.e. with
the little group of massive states (the rotation group) in 1+10 dimensions.
This was discovered in the process of analyzing D=11 supermembrane theory,
but the same arguments apply directly to type IIA string as is evident in
\cite{ibmembrane}. It will also be repeated below.

In this paper we will make a proposal that extends this observation to
include the higher stringy levels, but only after including non-perturbative
states. We will give a systematic prescription for the masses and SO(8) or
SO(9) quantum numbers of the non-perturbative stringy states, and will carry
out an explicit analysis of the D=11 structure successfully up to string
level 5. The perturbative states combined with the non-perturbative ones
will form complete supermultiplets at each string level. In the weak
coupling limit the non-perturbative states become infinitely heavy or
decouple, and the spectrum reduces to the usual type IIA perturbative string
states. In the infinite coupling limit there is a Kaluza-Klein tower of
degenerate massive states forming D=11 supermultiplets at each stringy level.

\section{Perturbative string spectrum}

In 10 dimensions (1- time + 9-space) the rotation group is SO(9). A massive
state at rest must come in degenerate multiplets of SO(9), where the
multiplet represents the spin components. When one actually constructs the
states of the type IIA superstring in the lightcone gauge there is only a
manifest SO(8) symmetry (see appendix). This is because manifest Lorentz
invariance is broken by the choice of gauge. However, since the theory is
actually Lorentz invariant one finds that the SO(8) representations can be
re-assembled into SO(9) representations (see e.g.\cite{GSW}).

Furthermore, there is separate supersymmetry for the left movers and right
movers (type IIA). For left movers, or right movers, there are 16
supercharges. For a massive state at rest $p^\mu =(M,\vec 0$) the
supercharges form a 16 dimensional Clifford algebra for left movers $\left\{
S_\alpha ^L,S_\beta ^L\right\} =M\delta _{\alpha \beta }$, and a similar one
for right movers. The automorphism group of this algebra is SO(16). It is
useful to embed SO(9) in this SO(16) by mapping the vector of SO(16) into
the 16 dimensional spinor of SO(9). The 16 supercharges may be rearranged
into 8 fermionic creation operators and 8 annihilation operators. At the
most 8 powers of the creation operators can be applied on a given state. The
repeated action of the supercharges can be organized into the two spinor
representations $2_B^7+2_F^7$ of SO(16), where the subscripts $B,F\,\,\,$%
imply that they are bosonic (even powers) or fermionic (odd powers)
operators respectively. The SO(9) content is obtained by decomposing these
SO(16) representations into the SO(9) representations
\begin{equation}
2_B^7=44+84,\quad 2_F^7=128
\end{equation}

Therefore, for either the left movers or right movers, the type IIA
superstring states must be arranged into massive supermultiplets of the form
\begin{equation}
r\times \{\left( 44+84\right) _B+128_F\}
\end{equation}
where $r$ is a representation of SO(9) that may be considered as the lowest
state in the supermultiplet (which may be bosonic or fermionic) and the
factor $\{\left( 44+84\right) _B+128_F\}$ represents the action of the
supercharges.

The closed string state is obtained by taking the direct product of left and
right movers at the same excitation level $l$, with $L_0=-M^2+l,\,\,\tilde L%
_0=-M^2+\tilde l$ and $L_0=\tilde L_0=0.$ The combined left and right moving
states at level $l$ take the form
\begin{equation}
\begin{array}{c}
\left( \sum_ir_i^{(l)}\right) _L\times \left( \sum_ir_i^{(l)}\right) _R \\
\times \left[
\begin{array}{l}
\{\left( 44+84\right) _B+128_F\}_L \\
\times \{\left( 44+84\right) _B+128_F\}_R
\end{array}
\right]
\end{array}
\label{perturb}
\end{equation}
where an identical collection of representations $\left(
\sum_ir_i^{(l)}\right) $ occur for left or right moving states at each level
$l$. These represent the lowest states in the supermultiplet.

Thus, to chracterize the states of the type IIA superstring it is sufficient
to give the collection of states $\sum_ir_i^{(l)}\,$ that occur at each
level $l$. Up to level 5 these are computed in the appendix, and are given
in the following table in terms of SO(9) representations
\begin{equation}
\begin{array}{ll}
\underline{Level\ } & \underline{SO(9)\ Representations\ \sum_ir_i^{(l)}} \\
l=1:\quad & 1_B \\
l=2: & 9_B \\
l=3: & 44_B+16_F \\
l=4: & \left\{
\begin{array}{l}
(9+36+156)_B \\
+128_F
\end{array}
\right. \\
l=5: & \left\{
\begin{array}{l}
\left( 1+36+44+84+231+450\right) _B \\
+\left[ 16+128+576\right] _F
\end{array}
\right.
\end{array}
\label{perturbative}
\end{equation}
Levels $l=0,1,2$ were previously given \cite{GSW} while the results for
levels $l=4,5$ are new. The SO(9) irreducible tensor structure of these
representations are given in (\ref{sugra}) and below
\begin{equation}
\begin{array}{l}
\phi _{(IJKL)}=450,\quad \phi _{(IJ,K)}=231,\quad \phi _{(IJK)}=156, \\
\phi _{[IJ]}=36,\quad {{{\psi _{\alpha (IJ)}=576.}}}
\end{array}
\end{equation}
Indices in square brackets $\phi _{[IJ]}$ are anti-symmetrized$,\,$ in round
parantheses $\phi _{(IJK)}$ etc. are completely symmetrized and traces
projected out, $\phi _{(IJ,K)}$ corresponds to a Young tableaux with (2,1)
boxes and traces projected out, while ${{{\psi _{\alpha (IJ)}}}}$ is a mixed
spinor-tensor with a projection involving a gamma matrix $[16\times
45-16\times 9=576].$

\section{D=11 supermultiplet structure}

\subsection{Perturbative states}

Type IIA superstring in D=10 has two supercharges of opposite chirality.
Each supercharge has 16 components and all states can be classified as SO(9)
supermultiplets as seen above. However, there is a higher supermultiplet
structure. To understand this, first notice that, at rest, the two
supercharges combined form a 32 dimensional Clifford algebra, which may be
divided into 16 creation operators and 16 annihilation operators. The
isomorphism group of this algebra is SO(32). The 16 creation operators form
the {\bf 16} dimensional representation of the rotation group SO(9).
However, it is useful to regard this SO(9) as being embedded in SO(32) as
follows
\begin{equation}
SO(32)\supset SU(16)\supset SO(10)\supset SO(9)
\end{equation}
where the embedding is done by classifying the 16 creation (annihilation)
operators in the {\bf 16} ($\overline{{\bf 16}}$) of SU(16), which is the
{\bf 16} ($\overline{{\bf 16}}$) of SO(10) and the {\bf 16} ({\bf 16}) of
SO(9). This embedding exhibits an intermediate SO(10) which will play an
essential role below. This is the embedding that we used some time ago\cite
{ibmembrane}.

To take advantage of this higher structure, let us reorganize the
perturbative levels as follows. A massive supermultiplet in the type IIA
superstring is obtained by starting from any representation of SO(9) (which
may represent either bosons or fermions) and applying the 16 fermionic
creation supercharges on it repeatedly. At the most 16 powers can be
applied. The SO(9) content of the $n^{th}$ power of the generator is
obtained by antisymmetrizing the $16$ of SO(9) $n$-times, i.e. $\left[
16^n\right] $. The SO(9) content of this antisymmetrization is better
understood by embedding it in SO(32), since the same procedure forms the two
spinor representations of SO(32). Thus, the even powers form the spinor $%
2_B^{15}$ and the odd powers form $2_F^{15}$ where the subscripts $B,F$
stand for bosons or fermions. Each spinor may be decomposed into
representations of SU(16). The $2_B^{15}$ contains the completely
antisymmetric SU(16) tensors with $0,2,4,6,8,10,12,14,16$ indices, and
likewise the $2_F^{15}$ contains the completely antisymmetric SU(16) tensors
with $1,3,5,7,9,11,13,15$ indices. Therefore, under SU(16) we have the
representations
\begin{eqnarray}
2_B^{15} &=&\left\{
\begin{array}{l}
\,\,\,\,\,1+120+1820+8008+12870 \\
+\overline{1}+\overline{120}+\overline{1820}+\overline{8008}
\end{array}
\right. \\
2_F^{15} &=&\left\{
\begin{array}{c}
\,\,\,\,16+560+4368+11440 \\
+\overline{16}+\overline{560}+\overline{4368}+\overline{11440}
\end{array}
\right.  \nonumber
\end{eqnarray}
These are further decomposed under SO(10) and then under SO(9)\cite
{ibmembrane}. Thus, any massive supermultiplet must have the structure of
the long D=11 supermultiplet
\begin{equation}
R\times \left\{ 2_B^{15}+2_F^{15}\right\}  \label{long}
\end{equation}
where $R$ is some collection of SO(9) representations (rather than SO(10) at
this stage) representing either a boson or fermion, and the structure $%
\left\{ 2_B^{15}+2_F^{15}\right\} $ comes from applying the supercharges on
it.

By comparing to the perturbative spectrum in eq.(\ref{perturb}) we can
determine that
\begin{equation}
\begin{array}{c}
\left\{ 2_B^{15}+2_F^{15}\right\} =\left\{
\begin{array}{l}
\{\left( 44+84\right) _B+128_F\}_L \\
\times \{\left( 44+84\right) _B+128_F\}_R
\end{array}
\right. \\
\\
{Perturbative: \quad }R=R_P^{(l)}\equiv \left( \sum_ir_i^{(l)}\right)
_L\times \left( \sum_ir_i^{(l)}\right) _R
\end{array}
\label{longg}
\end{equation}
The point is that the action of the supercharges represented by $\left\{
2_B^{15}+2_F^{15}\right\} $ has a higher symmetry structure. In particular
we focus on its SO(10) subgroup since it is the rotation group in 10
space-like dimensions. Indeed, from the D=11 the superalgebra point of view
the SO(10) has the correct interpretation to be associated with one
additional space-like dimension.

The question is whether the factor $R$ also has the SO(10) symmetry? The
answer is affirmative \cite{ibmembrane} for the first excited level $l=1$
since, according to (\ref{perturbative}) we have only SO(9) singlets
\[
l=1:\quad R_P^{(1)}=1_B\times 1_B=1,
\]
which are also SO(10) singlets! Thus, in addition to the arguments given by
Witten about the existence of hidden D=11 structure at the massless level,
we have a first hint that D=11 evidence may show at the massive stringy
levels as well. The argument above relies only on the supersymmetry
structure and is completely general as far as the first level is concerned.

At levels $l>1$ we need to analyse $R_P$ in more detail. For example at
level 2 we have from (\ref{longg},\ref{perturbative})
\begin{equation}
l=2:\quad R_P^{(2)}=9_B\times 9_B  \label{incomplete}
\end{equation}
which clearly cannot be complete SO(10) multiplets. So, there is no way that
the perturbative spectrum has D=11 structure by itself.

\subsection{Non-perturbative states}

We now postulate that the D=10 type-IIA superstring has additional
non-perturbative states that emerge just like monopoles or black holes in
non-linear theories. According to the supersymmetry algebra their mass
should satisfy $M\geq \frac c\lambda \left| W\right| .$ In the weak coupling
limit of the type IIA superstring these states are presumably infinitely
heavy, or infinitely weakly coupled (i.e. non-perturbative $e^{-c/\lambda })$%
, and therefore do not appear in the perturbative theory. In the strong
coupling limit, if there is D=11 structure, the extra non-perturbative
states should become degenerate with the perturbative states just in such a
way as to give complete SO(10) supermultiplets. In this case, since the D=11
momentum of the state has the form $P=(p^\mu ,cW/\lambda ),$ we conjecture a
stringy mass relation of the type
\begin{equation}
P^2=M_{11}^2=M^2-c^2W^2/\lambda ^2=n,
\end{equation}
where the stringy $n=0,1,2,\cdots $ is a positive integer.

The $n=0$ case is the sector that saturates the BPS bound, and therefore the
states form the short multiplets of D=11 supersymmetry with $%
2_B^7+2_F^7=(44+84)_B+128_F,$ just like the massless states, since for all
of them the D=11 mass vanishes, $P^2=M_{11}^2=0$. The elementary massless
states of the string have $W=0,$ but the non-perturbative states have $W\neq
0.$ The $n=0$ case is the sector discussed by Witten, and interpreted as
Kaluza-Klein modes of D=11 supergravity, with charges $W=kW_0,$ and $M=\frac
c\lambda \left| W\right| $.

In our case we investigate the conjectured stringy non-perturbative states,
with $n\geq 1.$ The masses will be $M=\sqrt{n+c^2W^2/\lambda ^2}.$ Just like
the $n=0$ case, it is natural to expect one state with $W=0,$ and an
infinite number of charged states with $W=kW_0.$ At infinite coupling $%
c^2W^2/\lambda ^2\rightarrow 0,$ all the new non-perturbative states of
level $n$ have the same mass $M^2=n$, which is similar to the spectrum of
the elementary massive string states. At weak coupling the $W\neq 0$ states
become infinitely heavy and disappear from the spectrum. But how about the $%
W=0,$ $M^2=n$ non-perturbative states which remain at finite mass at weak
coupling. Assuming that the $W=0$ states do not exist at all is one option,
but it seems unnatural, since $W=0$ is interpreted as just a zero value for
the 11th component of the momentum. Instead I will assume that its coupling
to the elementary string states is non-perturbative, e.g. $g\sim \exp
(-c/\lambda )$ rather than a power of $\lambda .$ This would explain why it
would not be seen in weakly coupled perturbation expansion of the type-IIA
superstring.

Our aim here is to answer the question ``what is the massive stringy
spectrum that exhibits D=11 structure and SO(10) symmetry at strong
coupling?''. The implication is that the extra non-perturbative states
together with the perturbative states form complete SO(10) supermultiplets.
However, can we provide a concrete formula for identifying all such states
at all massive levels? Here we make a proposal for all levels, and show that
it works at least up to level 5.

As a first hint consider how the $l=2$ states in (\ref{incomplete}) can be
completed to SO(10) multiplets. We need to add singlets for both left movers
and right movers at the same mass. By examining (\ref{perturbative}) we see
that the only place where singlets appear are at the previous level ($l=1).$
This suggests the presence of new non-perturbative states at level $l$ (with
$l=2\,$ in the present example), whose quantum numbers are those of the
perturbative states of the previous levels. Indeed this hint leads to the
following systematic formula.

In the infinitely strong coupling limit (or for the $W=0$ state) we suggest
that the mass formula for left and right movers that incorporates all
perturbative and non-perturbative states follow from putting the
non-perturbative string on shell as follows
\begin{eqnarray}
L_0 &=&-M^2+l^{\prime }+l=0, \\
\tilde L_0 &=&-M^2+\tilde l^{\prime }+\tilde l=0,  \nonumber
\end{eqnarray}
where $l^{\prime }$ is a positive integer and $l$ represents the level of
excitation with the elementary superstring oscilators as defined before.
Then integer $l$ determines the SO(9) content of the state as given in (\ref
{perturbative}). The factors $\left( -M^2+l^{\prime }\right) ,\,$ $(-M^2+%
\tilde l^{\prime })$ play the role of D=11 squared masses ($-M_{11}^2)$ for
left and right movers respectively. Thus, at a fixed 10D mass level $%
M^2=n=l^{\prime }+l$ the integers $\left( l^{\prime },l\right) ,(\tilde l%
^{\prime },\tilde l)$ take the values
\begin{eqnarray}
\left( l^{\prime },l\right) &=&\left( 0,n\right) +(1,n-1)+\cdots \left(
n-1,1\right) ,  \label{states} \\
(\tilde l^{\prime },\tilde l) &=&\left( 0,n\right) +(1,n-1)+\cdots \left(
n-1,1\right) ,  \nonumber
\end{eqnarray}
The states with the highest level $\left( l^{\prime }=0,l=n\right) $ are
isomorphic to the perturbative level $l=n$ states that were discussed in the
previous section. The others are the non-perturbative states with $l^{\prime
}\neq 0$. The SO(9) quantum number of the state $\left( l^{\prime },l\right)
$ is identical to those listed in eq.(\ref{perturbative}), as determined
only by $l,$ but the new states listed in (\ref{states}) are distinguished
from those in (\ref{perturbative}) by their quantum number $l^{\prime }$, as
well as the $W-$charge, if any.

Note that we have not included the state $(l^{\prime }=n,l=0)$ even though
the mass formula $M^2=l^{\prime }+l=n$ allows it. The reason is that the
D=11 left/right masses vanish when $l=\tilde l=0,$ i.e. $M_{11}^2=\left(
M^2-l^{\prime }\right) =(M^2-\tilde l^{\prime })=0$ and then the
supermultiplet is the short one (the BPS saturated states) $%
2_B^7+2_F^7=(44+84)_B+128_F.$ These states are stringy partners of the
Kaluza-Klein excitations of the D=11 supergravity multiplet. They may not
exist at all, or they may be interpreted as point-like states (from D=10
point of view) that would fit into the type of discussion given by Witten.
The stringy states must have $l\neq 0$ and must form the long multiplets $%
\left\{ 2_B^{15}+2_F^{15}\right\} \times R$ as described in (\ref{long},\ref
{longg}). Therefore, since $(l^{\prime }=n,l=0)$ are not such states they
are excluded from the stringy list in (\ref{states}).

In this section we simply want to use this mass formula irrespective of its
origin and show that it works. However, in addition to the mass formula we
need a scheme for the {\it multiplicity} of each state listed in (\ref
{states}). We will explore two schemes for the multiplicities. These will
differ from each other by how many times it is possible to obtain the same
value of $l^{\prime }.$ Each scheme works at least up to level 5, but each
one provides a different view on the origin of $l^{\prime }$ and the
dynamics of the weakly coupled dual theory. The simplest scheme is to take a
{\it single copy} of each state listed in (\ref{states}). We now show how
the SO(9) representations re-assemble to give SO(10) representations at each
mass level $M^2=n\geq 1$.

At mass level $M^2=1$ there are only the states $(0,1)$ that correspond to
the $2_B^{15}+2_F^{15}$ perturbative states already discussed above. They
form the D=11 supermultiplet
\begin{equation}
M^2=1:1\times \left\{ 2_B^{15}+2_F^{15}\right\} .
\end{equation}
This is SO(10) covariant.

At mass level $M^2=2$ we have for left movers
\begin{equation}
\left\{ \left( 0,2\right) +\left( 1,1\right) \right\} _L:\quad R=\left\{
9_B+1_B\right\} _L
\end{equation}
and the same set of representations for the right movers. These SO(9)
representations are read-off directly from eq.(\ref{perturbative}) through
their $l$-values . Therefore, the full set of perturbative and
non-perturbative states form the D=11 supermultiplet
\begin{equation}
M^2=2:\left( 10\times 10\right) \left\{ 2_B^{15}+2_F^{15}\right\} .
\end{equation}

At mass level $M^2=3$ we read off from eq.(\ref{perturbative}) the SO(9)
content of $(0,3)+\left( 1,2\right) +\left( 2,1\right) .$ For either left or
right movers this is $R=\left( 44+9+1\right) _B+16_F.$ These reassembe into
SO(10) representations so that the D=11 supermultiplet is
\begin{equation}
M^2=3:\left\{ 54_B+16_F\right\} \times \left\{ 54_B+\overline{16}_F\right\}
\times \left\{ 2_B^{15}+2_F^{15}\right\} .
\end{equation}
where $\phi _{(XY)}=54$ is the symmetric traceless tensor for SO(10) and $16,%
\overline{16}$ are the two spinor representations.

At mass level $M^2=4$ the SO(9) content of $(0,4)+(1,3)+\left( 2,2\right)
+\left( 3,1\right) $ is obtained from (\ref{perturbative}). For left movers
the SO(9)\ representations reassemble into SO(10) as follows
\begin{eqnarray}
SO(9) &=&\left\{
\begin{array}{l}
(1+2\times 9+36+156+44)_B \\
+\left( 16+128\right) _F
\end{array}
\right.  \label{l49} \\
SO(10) &=&\left( 45+210\right) _B+144_F.  \label{l410}
\end{eqnarray}
Therefore the full set of perturbative and non-perturbative states included
in the D=11 supermultiplet is
\begin{equation}
M^2=4:\left\{
\begin{array}{c}
\left\{ \left( 45+210\right) _B+144_F\right\} \\
\times \left\{ \left( 45+210\right) _B+\overline{144}_F\right\} \\
\times \left\{ 2_B^{15}+2_F^{15}\right\}
\end{array}
\right.
\end{equation}
Here the SO(10) tensors are
\begin{eqnarray}
\phi _{(XYZ)} &=&210,\quad \phi _{[XY]}=45, \\
\psi _{\alpha X} &=&144,\quad \psi _{\bar \alpha X}=\overline{144}.
\nonumber
\end{eqnarray}
and their SO(10)$\rightarrow $SO(9) decomposition is
\begin{eqnarray}
210 &\rightarrow &156+44+9+1 \\
45 &\rightarrow &36+9  \nonumber \\
144 &\rightarrow &128+16  \nonumber
\end{eqnarray}

Finally, at mass level $M^2=5$ the SO(9) content of $(0,5)+(1,4)+(2,3)+%
\left( 3,2\right) +\left( 4,1\right) $ for left movers, as obtained from (%
\ref{perturbative}), reassemble into SO(10) representations as follows
\begin{eqnarray}
SO(9) &=&\left\{
\begin{array}{l}
\left(
\begin{array}{c}
2\times \left( 1+9+36+44\right) \\
+84+156+231+450
\end{array}
\right) _B \\
+\left[ 2\times \left( 16+128\right) +576\right] _F
\end{array}
\right.  \label{l59} \\
SO(10) &=&\left\{
\begin{array}{c}
\left( 1+120+320+660\right) _B \\
+\left[ 144+720\right] _F
\end{array}
\right.  \label{l510}
\end{eqnarray}
Therefore the full set of perturbative and non-perturbative states forming
the D=11 supermultiplet is
\begin{equation}
M^2=5:\left\{
\begin{array}{c}
\left\{
\begin{array}{c}
\left( 1+120+320+660\right) _B \\
+\left[ 144+720\right] _F
\end{array}
\right\} \\
\times \left\{
\begin{array}{c}
\left( 1+120+320+660\right) _B \\
+\left[ \overline{144}+\overline{720}\right] _F
\end{array}
\right\} \\
\times \left\{ 2_B^{15}+2_F^{15}\right\}
\end{array}
\right.
\end{equation}
The SO(10) tensors are

\begin{eqnarray}
\phi _{(UXYZ}) &=&660,\quad \phi _{(XY,Z)}=320,\quad \phi _{[XYZ]}=120
\nonumber \\
\psi _{\alpha (XY)} &=&{{720,\quad \psi _{\alpha X}=144}} \\
\psi _{\bar \alpha (XY)} &=&\overline{{720}}{{,\quad \psi _{\bar \alpha X}=}}%
\overline{{144}}  \nonumber
\end{eqnarray}
Their SO(10)$\rightarrow SO(9)$ decomposition is
\begin{eqnarray}
660 &\rightarrow &450+156+54+9+1  \nonumber \\
320 &\rightarrow &231+44+36+9  \nonumber \\
120 &\rightarrow &84+36 \\
720 &\rightarrow &576+128+16  \nonumber \\
144 &\rightarrow &128+16.  \nonumber
\end{eqnarray}

\subsection{Higher multiplicity scheme}

The simplest scheme for the multiplicity of the states that was used above
may be effectively reformulated as follows. Introduce a new left-moving
oscillator $\beta _{-1}$ that has a single mode at level 1. A similar one $%
\tilde \beta _{-1}$ is introduced also for right movers. Then we may
identify the number operators $l^{\prime }=\beta _{-1}\beta _1$ and $\tilde l%
^{\prime }=\beta _{-1}\beta _1$ while the states $\left( l^{\prime
},l\right) ,(\tilde l^{\prime },\tilde l)$ can be built as as follows
\begin{eqnarray}
\left( l^{\prime },l\right) &\longleftrightarrow &\left( \beta _{-1}\right)
^{l^{\prime }}|l>_L, \\
(\tilde l^{\prime },\tilde l) &\longleftrightarrow &\left( \tilde \beta
_{-1}\right) ^{\tilde l^{\prime }}|\tilde l>_R,  \nonumber
\end{eqnarray}
where $|l>_L,|\tilde l>_R$ are the original string oscillator states given
in (\ref{perturbative}). This construction should be regarded as a mnemonic
to keep track of the states, and while it is not excluded, it need not
represent necessarily an additional oscillator in the dual theory. This
minimal approach clearly gives a single copy of each state $\left( l^{\prime
},l\right) ,(\tilde l^{\prime },\tilde l).$

It is tempting to explore the idea of extra oscillators, leading to higher
multiplicities. The most attractive case would be to boldly introduce all
modes for an extra dimension. This would account naturally for all stringy
Kaluza-Klein partners in the conjectured D=11 dual theory. Let the modes be $%
\beta _n$ with the usual Heisenberg algebra $[\beta _n,\beta _m]=n\delta
_{n+m}.$ Similarly introduce also right movers $\tilde \beta _n.$ The zero
mode is the Kaluza-Klein charge that appears as the 11th momentum in the
superalgebra $\beta _0=\tilde \beta _0\sim W.$ We may now construct many
more states with the same value of $l^{\prime },\tilde l^{\prime }$%
\begin{equation}
l^{\prime }=\sum_n\beta _{-n}\beta _n,\quad \tilde l^{\prime }=\sum_n\tilde
\beta _{-n}\tilde \beta _n
\end{equation}
At level $l^{\prime }$ the states $(l^{\prime },l)$ are
\begin{equation}
\beta _{-l^{\prime }}|l>_L\oplus \beta _{-l^{\prime }+1}\beta
_{-1}|l>_L\oplus \cdots \oplus \left( \beta _{-1}\right) ^{l^{\prime }}|l>_L
\end{equation}
The multiplicity is equal to the number of partitions $p(l^{\prime })$ of
the integer $l^{\prime }.$ Thus, in this scheme we would have the following
states up to mass level $M^2=5$%
\begin{equation}
\begin{array}{ll}
\,\underline{M^2=n} & \,\quad \underline{States\,\,\sum \,p(l^{\prime
})\times (l^{\prime },l)} \\
1 & (0,1) \\
2 & (0,2)+(1,1) \\
3 & (0,3)+(1,2)+2(2,1) \\
4 & (0,4)+(1,3)+2(2,2)+3(3,1) \\
5 & (0,5)+(1,4)+2(2,3)+3(3,2)+5(4,1)
\end{array}
\label{mul}
\end{equation}
The remarkable fact is that these states also can be reorganized into
complete SO(10) supermultiplets. This is done as follows.

There is nothing new at mass levels $M^2=1,2.$ At mass level $M^2=3,$ first
use one factor of each state to obtain the result $\left\{ 54_B+16_F\right\}
$ as before. The remaining $(2,1)$ is just a singlet of SO(10); that is, it
has the same SO(10) content as mass level $n=1$. Therefore, the full
collection of SO(10) states is $\left\{ \left( 54+1\right) _B+16_F\right\} $
for left movers, and the same one for the right movers. These multiply the
overall factor $2_B^{15}+2_F^{15}$ as before.

Similarly, at mass level $M^2=4$, after reproducing the previous collection
of SO(10) states in (\ref{l410}) by using one factor of each state, there
remains $(2,2)+2(3,1).$ This may be regarded as two sets $(2,2)+(3,1)$ and $%
(3,1)$ each having multiplicities one, and furthermore having the same SO(9)
or SO(10) content as the collection of all the states in mass levels $n=1,2$
as listed in (\ref{mul})$.$ Using the known result for those levels, we see
that the extra states correspond to the SO(10) multiplets $\left(
10+1\right) _B.$ Therefore the full collection of left-moving SO(10) states
is $\left( 1+10+45+210\right) _B+\left[ 16+144\right] _F.$ Similarly, for
right movers.

Finally at mass level $M^2=5,$ the same procedure indicates that in addition
to the states in (\ref{l510}) there are those coming from the levels $%
(2,3)+2(3,2)+4(4,1).$ Again, these may be regarded as several sets, each
containing single multiplicities,
\begin{eqnarray}
&&\ (2,3)+(3,2)+(4,1)  \nonumber \\
&&\ (3,2)+(4,1) \\
&&\ (4,1)  \nonumber \\
&&\ (4,1)  \nonumber
\end{eqnarray}
These have the same SO(9) or SO(10) quantum numbers computed for the single
multiplicity scheme at the lower mass levels $n=1,2,3.$ Therefore the total
SO(10) multiplets are
\begin{eqnarray}
&&\ \left( 5\times 1+2\times 10+54+120+320+660\right) _B \\
&&\ +\left[ 16+144+720\right] _F  \nonumber
\end{eqnarray}

The pattern is clear. If we have already established that the single
multiplicity scheme gives SO(10) multiplets, then the expanded scheme also
gives it since the additional states at mass level $M^2=n$ have the same
SO(10) quantum numbers as the states in the mass levels $M^2=(n-2),(n-3),%
\cdots ,1.$ We have already shown this up to level 5, and an iterative proof
can be given for all mass levels.

Recall that in addition to these stringy Kaluza-Klein partners, the dual
theory presumably has also an infinite tower of point-like Kaluza-Klein
states for all possible values of the 11th momentum $W$ (all of which become
degenerate at infinite coupling).

\section{Comments}

I have made a proposal for identifying the conjectured non-perturbative
stringy states that uncover a hidden D=11 structure. This involves stringy
structures that are not included in Witten's discussion, but which must be
there if his proposal is more than an accident at low energies. In order to
tighten the arguments one should look for a possible role of discrete
symmetries, similar to, or beyond the conjectured SL(2,Z) symmetry and the
associated U-duality \cite{hulltowns}. There may be a symmetry that commutes
with the SO(8) or even SO(9), but not with SO(10). Combining such a symmetry
with SO(10) may generate a much more restrictive symmetry of the string
theory that could be sufficient to elucidate the conjectured
eleven-dimensional structure and the strong coupling behavior of the theory.
Evidence for such a symmetry would begin with finding repetitions of the
same SO(8) or SO(9) representations at the same mass levels, such that they
would form multiplets of the extra symmetry. We indeed find such repetitions
at various stages of our analysis as is evident in the appendix and the
text. However, more is needed to understand if this is due to a symmetry.

It may be that restoring the lightcone oscillators would make it easier to
investigate the presence of symmetries in a covariant quantization. In this
connection I suspect that a new construction of an SL(2,R) current algebra
that uses the lightcone oscillators \cite{ibsl2wzw} would be useful. Note
that the arbitrary $c=0$ stress tensor $T^{\prime }$ used in this
construction may be taken as the stress tensor of the type-IIA theory by
including ghost fields. Extra effective dimensions may arise through such a
mechanism.

One approach for searching for the non-perturbative states is to investigate
a string field theory type formulation. In this connection recent proposals
for the field theoretic formulation of the superstring by Berkovits and Vafa
seem promising\cite{bv}.

The extra oscillators $\beta _n,\tilde \beta _n$ are intriguing. Is it
possible to include them directly in the discussion of the D=10 type-IIA
superstring, and not only in the dual theory? In this connection, perhaps it
is useful to recall that the covariant quantization of the Green-Schwarz
string has never been fully understood. Could there be a possibility of a
Liouville-like mode that decouples perturbatively, but which is present
non-perturbatively? If so, it would account naturally for the extra
dimension.

The dual D=11 theory that we are seeking, especially when discussed in terms
of the oscillators $\beta _n,\tilde \beta _n,$ is beginning to look like a
membrane theory, as it shares some similar features to the D=11
supermembrane theory that we studied some time ago, although not quite the
same. The D=11 theory that is dual to type-IIA superstring may be a new
membrane-like theory. In any case, in view of the approach we have pursued
here, it may be useful to revise also the previous work \cite{ibmembrane}%
\cite{bapose}\cite{ibanomaly}, using the new hints as a guide for the
construction of a consistent ``supermembrane theory'' in 11 dimensions.

\section{Acknowledgements}

I thank K. Pilch and J. Minahan for discussions, and E. Witten and J.
Schwarz for comments.

\section{Appendix}

In this appendix we construct the SO(9) content of the elementary string
states for D=10 type-IIA superstring, up to level 5. The results were
previously known for levels 0,1,2,3 \cite{GSW}, but they are new for levels
4 and 5.

\subsection{Notation}

We work directly in the light-cone gauge. Therefore, we start with manifest
SO(8) symmetry. Vector indices $i=1,2,\cdots 8$ denote the vector $8_v$ of
SO(8), while the spinor indices $a,\dot a=1,2,\cdots 8,$ denote the spinor
representations $8_{+},8_{-}$ respectively. The string oscillators are
classified as $\alpha _{-n}^i=8_v,\,S_{-n}^a=8_{+}$ for left movers and $%
\tilde \alpha _{-n}^i=8_v,\,\,\tilde S_{-n}^{\dot a}=8_{-}$ for right
movers. The zero modes $S_0^a=8_{+},\,\,\tilde S_0^{\dot a}=8_{-}$ give the
Ramond vacuum $|8_v+8_{-}>_L$ for left movers and $|8_v+8_{+}>_R$ for right
movers.

SO(9) vector indices are denoted by $I=1,2,\cdots 9$ and spinor indices by $%
\alpha =1,2,\cdots 16.$ SO(10) vector indices are denoted $X=1,2,\cdots 10$
and spinor indices by $\alpha ,\bar \alpha =1,2,\cdots 16$ for the spinor
representations $16,\overline{16}$ of SO(10) respectively. The
group/subgroup decomposition of these representations is

\begin{equation}
\begin{array}{ccc}
SO(10) & SO(9)
\begin{array}{l}
\\
\end{array}
& SO(8) \\
\phi _X=10 & \left\{
\begin{array}{l}
\phi _I=9 \\
\phi =1
\end{array}
\right. & \left\{
\begin{array}{l}
\left( \phi _i=8_v\right) +\left( \phi ^{\prime }=1\right) \\
+\left( \phi =1\right)
\end{array}
\right. \\
\psi _\alpha =16 & \psi _\alpha =16 & \left( \psi _a=8_{+}\right) +\left(
\bar \psi _{\dot a}=8_{-}\right) \\
\bar \psi _{\bar \alpha }=\overline{16} & \psi _\alpha ^{\prime }=16 &
\left( \psi _a^{\prime }=8_{+}\right) +\left( \bar \psi _{\dot a}^{\prime
}=8_{-}\right)
\end{array}
\end{equation}
Thus, the decomposision is obtained by specializing the index $X\rightarrow
I\oplus 10\rightarrow i\oplus 9\oplus 10$ and $\alpha ,\bar \alpha
\rightarrow \alpha \rightarrow a\oplus \dot a$ .

For SO(n) all antisymmetric tensors are irreducible multiplets, but tensors
with symmetric vector indices or mixed spinor-vector indices are reducible.
Irreducible multiplets are obtained provided these tensors give zero when
contracted with the Kr\"onecker delta function or with a gamma matrix. Thus,
for SO(8), SO(9) or SO(10) writing irreducible representations in the form $%
\phi _{\left( XY\right) }.,$ or $\psi _{\alpha X}$ , etc. implies that these
tensors are constrained as follows
\begin{equation}
\phi _{(XY)}\,\delta ^{XY}=0,\quad \left( \gamma ^X\right) _\alpha ^\beta
\,\,\psi _{\beta X}=0.
\end{equation}
Combining these rules with the decomposition of indices in the above table,
one can figure out the group/subgroup decomposition of higher
representations. For example
\begin{equation}
\begin{array}{lll}
SO(10) &
\begin{array}{l}
\\
\end{array}
SO(9) & SO(8) \\
\phi _{(XY)} & \left\{
\begin{array}{l}
\phi _{(IJ)}+ \\
\phi _{I\,or\,J}+\phi
\end{array}
\right. & \left\{
\begin{array}{l}
(\phi _{(ij)}+\phi _{i\,or\,j}+\phi ^{\prime }) \\
+(\phi _{i\,or\,j}+\phi ^{\prime \prime })+\phi
\end{array}
\right. \\
=54 & =44+9+1 & \left\{
\begin{array}{l}
=(35_v+8_v+1) \\
+(8_v+1)+1
\end{array}
\right. \\
\psi _{\alpha X} & \psi _{\alpha I}+\psi _\alpha ^{\prime } & \left\{
\begin{array}{l}
\psi _{ai}+\psi _{ai} \\
+\psi _{\dot ai}+\psi _{\dot ai} \\
+(\psi _a+\psi _{\dot a}) \\
+(\psi _a^{\prime }+\psi _{\dot a}^{\prime })
\end{array}
\right. \\
=144 & 128+16 & \left\{
\begin{array}{l}
=\left(
\begin{array}{c}
56_{+}+56_{-} \\
+8_{+}+8_{-}
\end{array}
\right) \\
\quad +(8_{+}+8_{-})
\end{array}
\right.
\end{array}
\end{equation}

\subsection{Massless sector}

The Ramond-Ramond vacuum $\left( l=0\right) $ has the following SO(8)
classification of bosons and fermions.
\begin{eqnarray}
\begin{array}{l}
|8_v+8_{-}>_L\times |8_v+8_{+}>_R \\
=(8_v\times 8_v+8_{-}\times 8_{+})_B+(8_v\times 8_{+}+8_v\times 8_{-})_F \\
=(1+35_v+28+56_v+8_v)_B \\
\quad \quad +(56_{+}+8_{-}+56_{-}+8_{+})_F \\
{SO(9):\quad }=(44+84)_B+128_F\quad
\end{array}
\label{vacuum}
\end{eqnarray}
So, SO(9) emerges only after combining left with right.

\subsection{Level 1}

At massive levels, the left movers and right movers separately must exhibit
the SO(9) structure since each sector behaves like the open string. Another
reason, based on the representations of supersymmetry, was given in the
text. Therefore, we will first reorganize the left-moving SO(8) states into
SO(9) long supermultiplets ($2_B^7+2_F^7)\times r$ . The right moving states
have the identical SO(9) structure. The SO(10) structure will become
apparent only when left and right are put together, as done in the text.

At level 1, there are left and right moving oscillators applied on the left
and right vacuum. For left movers the SO(9) structure can be seen by writing
out the SO(8) content of the oscillators and vacuum
\begin{eqnarray}
{{{{Left}}}} &:&{{{(\alpha _{-1}^i\oplus S_{-1}^a)|{vac}>}}}_L
\label{left1} \\
&=&{{{\left( 8_v+8_{+}\right) \times \left( 8_v+8_{-}\right) }}}  \nonumber
\\
SO(9) &=&{\left[ (44+84)_B+128_F\right] _L}  \nonumber
\end{eqnarray}

\subsection{Level 2}

The oscillators applied on the left-vacuum are classified under SO(8) as
follows
\begin{eqnarray}
{Left} &:&\left[ (\alpha _{-2}^i\oplus S_{-2}^a)\oplus \left( (\alpha
_{-1}^i\oplus S_{-1}^a)^2\right) _{susy}\right] |{vac}>_L  \label{left2}
\\
\  &=&\left[ (8_v+8_{+})+\left( (8_v+8_{+})^2\right) _{susy}\right] \times
(8_v+8_{-})  \nonumber
\end{eqnarray}
where the subscript ``susy'' means symmetrization of identical bosons and
antisymmetrization of identical fermions. It can be rewritten as follows
\begin{eqnarray}
\left( (8_v+8_{+})^2\right) _{susy} &=&\left(
\begin{array}{l}
\left( 8_v\times 8_v\right) _S+\left( 8_{+}\times 8_{+}\right) _A \\
+\left( 8_v\times 8_{+}\right)
\end{array}
\right)   \nonumber \\
\  &=&\left( 1+35_v+28\right) _B+\left( 8_{-}+56_{+}\right) _F
\label{similar} \\
\  &=&8_v\times \left( 8_v+8_{+}\right)   \nonumber
\end{eqnarray}
where the subscripts $S,A$ mean symmetrization and antisymmetrization
respectively. The important last step is the rewriting in terms of an
overall factor $\left( 8_v+8_{+}\right) .$ This allows rewriting all the
SO(8) states for the left movers in (\ref{left2}) in the form of SO(9)
multiplets
\begin{eqnarray}
{{{{Left }}}} &:&{{{\left( 1+8_v\right) \times \left( 8_v+8_{+}\right)
\times \left( 8_v+8_{-}\right) }}} \\
\  &=&{9\times \left[ \left( 44+84\right) _B+128_F\right] }  \nonumber
\end{eqnarray}

\subsection{Level 3}

The oscillators that are applied on the vacuum have the SO(8) structure
\begin{eqnarray}
{Left} &:&\left[
\begin{array}{c}
(\alpha _{-3}^i\oplus S_{-3}^a) \\
\oplus (\alpha _{-2}^i\oplus S_{-2}^a)(\alpha _{-1}^i\oplus S_{-1}^a) \\
\oplus \left( (\alpha _{-1}^i\oplus S_{-1}^a)^3\right) _{susy}
\end{array}
\right] |{vac}>_L \\
\  &=&\left[
\begin{array}{c}
(8_v+8_{+})+(8_v+8_{+})^2 \\
+\left( (8_v+8_{+})^3\right) _{susy}
\end{array}
\right] \times (8_v+8_{-})  \nonumber
\end{eqnarray}
By an analysis similar to (\ref{similar}) we can rewrite the cubic
super-symmetrized product by factoring out $\left( 8_v+8_{+}\right) $ as
follows
\begin{equation}
\left( \left( 8_v+8_{+}\right) ^3\right) _{susy}=\left( 35_v+8_{-}\right)
\times \left( 8_v+8_{+}\right)   \label{cube}
\end{equation}
Then the left-moving states may be rewritten as SO(9) multiplets as follows
\begin{eqnarray}
{Left} &:&\left\{
\begin{array}{c}
\left( 1+\left( 8_v+8_{+}\right) +\left( 35_v+8_{-}\right) \right)  \\
\times \left( 8_v+8_{+}\right) \times \left( 8_v+8_{-}\right)
\end{array}
\right\}  \\
SO(9) &=&\left[ 44_B+16_F\right] \times \left[ \left( 44+84\right)
_B+128_F\right]   \nonumber
\end{eqnarray}
The SO(9) $\rightarrow $ SO(8) decomposition is
\begin{equation}
44\rightarrow 1+8_v+35_v
\end{equation}

\subsection{Level 4}

The oscillators that are applied on the vacuum have the SO(8) structure
\begin{eqnarray}
{Left} &:&\left[
\begin{array}{c}
(\alpha _{-4}^i\oplus S_{-4}^a) \\
\oplus (\alpha _{-3}^i\oplus S_{-3}^a)(\alpha _{-1}^i\oplus S_{-1}^a) \\
\oplus \left( (\alpha _{-2}^i\oplus S_{-2}^a)^2\right) _{susy} \\
\oplus (\alpha _{-2}^i\oplus S_{-2}^a)\left( (\alpha _{-1}^i\oplus
S_{-1}^a)^2\right) _{susy} \\
\oplus \left( (\alpha _{-1}^i\oplus S_{-1}^a)^4\right) _{susy}
\end{array}
\right] |{vac}>_L  \nonumber  \label{left4} \\
&=&\left[
\begin{array}{c}
(8_v+8_{+})+(8_v+8_{+})^2 \\
+\left( (8_v+8_{+})^2\right) _{susy} \\
(8_v+8_{+})\times \left( (8_v+8_{+})^2\right) _{susy} \\
+\left( (8_v+8_{+})^4\right) _{susy}
\end{array}
\right] \times (8_v+8_{-})  \label{left4}
\end{eqnarray}
First we work on rewriting the super-symmetrized quartic by pulling out a
factor of $(8_v+8_{+})$ as follows
\begin{equation}
\left( (8_v+8_{+})^4\right) _{susy}=\left[ \left( 8_v+112\right)
+56_{-}\right] \times (8_v+8_{+})  \label{quartic}
\end{equation}
where $\phi _{(ijk)}=112$ is the completely symmetric traceless SO(8) tensor
in three indices. Combining this result with (\ref{similar}) and inserting
them in (\ref{left4}) we may pull out the overall $(8_v+8_{+})$ factor to
exhibit the SO(9) classification as follows
\begin{eqnarray}
{Left} &:&\left\{
\begin{array}{l}
1+(8_v+8_{+})+8_v \\
+8_v\times (8_v+8_{+}) \\
+\left( 8_v+112\right) +56_{-}
\end{array}
\right\}   \nonumber \\
&&\ \quad \times (8_v+8_{+})\times (8_v+8_{-}) \\
\  &=&\left\{
\begin{array}{l}
\left(
\begin{array}{c}
2\times 1+3\times 8_v \\
+28+35_v+112
\end{array}
\right) _B \\
+\left( 8_{+}+8_{-}+56_{+}+56_{-}\right) _F
\end{array}
\right\}   \nonumber \\
&&\ \quad \quad \times (8_v+8_{+})\times (8_v+8_{-})  \nonumber \\
SO(9) &=&\left\{ (9+36+156)_B+128_F\right\}   \nonumber \\
&&\ \quad \times \left[ \left( 44+84\right) _B+128_F\right]   \nonumber
\end{eqnarray}
where $\phi _{(IJK)}=156$ and $\phi _{[IJ]}=36$ are the 3-index traceless
symmetric and the 2-index antisymmetric SO(9) tensors respectively. The
SO(9) $\rightarrow $ SO(8) decomposition is
\begin{eqnarray}
36 &\rightarrow &8_v+28 \\
156 &\rightarrow &1+8_v+35_v+112  \nonumber \\
128 &\rightarrow &8_{+}+8_{-}+56_{+}+56_{-}  \nonumber
\end{eqnarray}

\subsection{Level 5}

The oscillators that are applied on the vacuum have the SO(8) structure
\begin{eqnarray}
{Left} &:&\left[
\begin{array}{c}
(\alpha _{-5}^i\oplus S_{-5}^a) \\
\oplus (\alpha _{-4}^i\oplus S_{-4}^a)(\alpha _{-1}^i\oplus S_{-1}^a) \\
\oplus (\alpha _{-3}^i\oplus S_{-3}^a)(\alpha _{-2}^i\oplus S_{-2}^a) \\
\oplus (\alpha _{-3}^i\oplus S_{-3}^a)\left( (\alpha _{-1}^i\oplus
S_{-1}^a)^2\right) _{susy} \\
\oplus (\alpha _{-1}^i\oplus S_{-1}^a)\left( (\alpha _{-2}^i\oplus
S_{-2}^a)^2\right) _{susy} \\
\oplus (\alpha _{-2}^i\oplus S_{-2}^a)\left( (\alpha _{-1}^i\oplus
S_{-1}^a)^3\right) _{susy} \\
\oplus \left( (\alpha _{-1}^i\oplus S_{-1}^a)^5\right) _{susy}
\end{array}
\right] |{vac}>_L  \nonumber  \label{left5} \\
&&  \label{left5} \\
\  &=&\left[
\begin{array}{c}
(8_v+8_{+})+2\times (8_v+8_{+})^2 \\
+2\times (8_v+8_{+})\times \left( (8_v+8_{+})^2\right) _{susy} \\
+(8_v+8_{+})\times \left( (8_v+8_{+})^3\right) _{susy} \\
+\left( (8_v+8_{+})^5\right) _{susy}
\end{array}
\right] \times (8_v+8_{-})  \nonumber  \label{left5}
\end{eqnarray}
First we work on rewriting the super-symmetrized quintic by pulling out a
factor of $(8_v+8_{+})$ as follows

\begin{equation}
\left( (8_v+8_{+})^5\right) _{susy}=\left[
\begin{array}{c}
\left( 28+35+294\right) _B \\
+\left( 224_{-}+8_{-}\right) _F
\end{array}
\right] \times (8_v+8_{+})  \label{quintic}
\end{equation}
where
\begin{equation}
\phi _{(ijkl)}=294,\quad \psi _{\dot a(ij)}=224_{-}
\end{equation}
are the SO(8) representations. Combining this result with (\ref{similar},\ref
{cube}) and inserting them in (\ref{left5}) we may pull out the overall $%
(8_v+8_{+})$ factor and then exhibit the SO(9) classification as follows
\begin{eqnarray}
{Left} &:&\left\{
\begin{array}{c}
{1+2\times (8}_v{+8_{+})+2\times 8}_v{\times (8}_v{+8_{+})} \\
+{(35}_v{+8_{-})\times (8}_v{+8_{+})} \\
+{(28+35{+}294)+}\left( {224}_{-}+8_{-}\right)
\end{array}
\right\}   \nonumber \\
&&\ \ \quad \quad \quad \times \left\{ (8_v+8_{+})\times (8_v+8_{-})\right\}
\\
\  &=&\left\{
\begin{array}{c}
\left(
\begin{array}{c}
3{\times 1+4\times 8+3\times 28+3\times 35} \\
{+56+112+160+294}
\end{array}
\right) _B \\
+\left[
\begin{array}{c}
3\times 8_{+}+3\times 8_{-}+2\times 56_{+} \\
+2\times 56_{-}+224_{+}+224_{-}
\end{array}
\right] _F
\end{array}
\right\}   \nonumber \\
&&\ \ \quad \quad \quad \times \left\{ (8_v+8_{+})\times (8_v+8_{-})\right\}
\nonumber
\end{eqnarray}
These form the SO(9) multiplets
\begin{eqnarray}
SO(9) &=&\left\{
\begin{array}{c}
\left( 450+231+84+44+36+1\right) _B \\
+\left[ 576+128+16\right] _F
\end{array}
\right\}  \\
&&\ \ \quad \quad \times \left\{ \left( 44+84\right) _B+128_F\right\}
\nonumber
\end{eqnarray}
where $\phi _{(IJKL)}=450,$ $\phi _{(IJ,K)}=231,\,\,\phi
_{[IJK]}=84,\,\,\psi _{\alpha (IJ)}=576$ are the SO(9) irreducible tensors.
Their SO(9) $\rightarrow $ SO(8) decomposition is
\begin{eqnarray}
84 &\rightarrow &28+56  \nonumber \\
231 &\rightarrow &8+28+35+160 \\
450 &\rightarrow &1+8+35+112+294  \nonumber \\
576 &\rightarrow &8_{+}+8_{-}+56_{+}+56_{-}+224_{+}+224_{-}  \nonumber
\end{eqnarray}

\end{document}